# PREPRINT

# 13

# Incorporation of Journalistic Approaches Into Algorithm Design

Mariella Bastian, Damian Trilling, Mykola Makhortykh

# Abstract

The growing adoption of algorithm-powered tools in journalism enables new possibilities and raises many concerns. One way of addressing these concerns is by integrating journalistic practices and values into the design of algorithms that facilitate different journalistic tasks, from automated content generation to news content distribution. In this chapter, we discuss how such integration can happen. To do this, we first introduce the concepts of algorithms and different perspectives on algorithm design and then scrutinize various journalistic viewpoints on the matter and methodological approaches for studying these perspectives and their translation into specific algorithm-powered journalistic tools. We conclude by discussing important directions for future research, ranging from contextualizing journalistic approaches to algorithm design to accounting for the transformative impacts of artificial intelligence (AI) technologies.

# Introduction

The increasing adoption of algorithm-powered tools into journalistic newsrooms creates new possibilities but also risks. With the rise of online platforms, we have witnessed multiple challenges to traditional journalistic business models, practices, and values (see Chapter 4). The use of algorithm-powered tools exacerbates these challenges: Often, these tools are designed by software developers who do not always have extensive experience in journalistic work, and the workings of these tools remain obscure. As we will show in this chapter, these shortcomings are particularly significant as the design of new tools and their respective algorithms affect the whole chain of journalism, from news production to dissemination.

Deploying algorithm-powered tools into different sectors – including journalism – impacts people's daily lives in multiple ways. Some examples are obvious to many people, and some tools have become part of their daily lives. However, using these tools often happens in obscure ways, with people not being aware of algorithms being involved. In this chapter, we use the term "algorithm-powered tool" for a broad spectrum of applications, ranging from simple ones like ordering articles by the number of clicks to more advanced processes such as text generation using techniques often summarized under the umbrella term of artificial intelligence (AI) (see Chapters 1 and 2).

A prominent domain for the use of algorithm-powered tools is information dissemination. With the growing volume of available news due to 24/7 coverage of different societal processes, from elections to armed conflicts, individual information diets are increasingly shaped by algorithms. Though users might not always be fully aware of it, the design of algorithm-powered tools used by journalists has direct implications for the core aspects of democratic societies, including freedom of expression (Helberger et al., 2020) or the right to receive information (Eskens et al., 2017).

Under these circumstances, preventing algorithm-powered tools from undermining journalistic practices and cornerstones becomes paramount. To solve this issue, different approaches have to go hand in hand. It is important to investigate how journalists use algorithm-powered tools in their daily work and examine the user perspective on algorithms in the context of journalism. However, it is also crucial to consider how professional values and normative ideas of journalism can be incorporated into algorithm design. To prevent algorithms from undermining societal functions of journalistic media, journalistic approaches have to be integrated into algorithms powering specific tools from the beginning – and not integrated into the existing tools *post hoc* to address potential concerns their use may raise.

In this chapter, we aim to clarify how algorithm design affects journalism and how journalistic approaches can be supported by how the algorithms are designed. The chapter proceeds as follows: First, we outline the working of common algorithm-powered tools used in journalistic contexts. Then, we discuss journalistic perspectives on the algorithm design and how it affects different aspects of journalism, from journalists' roles and routines to news organizations, media markets, and the relationship to the audience. Finally, we look at different methodological approaches used to study algorithm design in journalism and conclude with a summary of our findings and the outlook.

## Design of Algorithms and Algorithm-Powered Tools

Algorithms are formalized descriptions of computational procedures (Dourish, 2016). By structuring the processes of computation and how these processes are developed, algorithms contribute to different sociotechnical assemblages, including those dealing with news production and distribution. While not synonymous with automation, algorithms are integral to the functionality of automated decision-making systems, including those increasingly used in journalism. Yet, even very simple computational procedures – like ranking articles by the number of clicks received from the readers – qualify as algorithms.

Algorithms used by journalists vary substantially in their functionality and composition, but societal and academic debates often focus on algorithms dealing with news distribution. One example is news recommender algorithms, which are computational procedures used for identifying news content that can be of particular interest to individual users. The specific implementations of news recommender algorithms vary (Karimi et al., 2018; Möller et al., 2018), but broadly, they can be categorized into four groups: (1) popularity-based, (2) collaborative filtering-based, (3) semantic-filtering-based, and (4) hybrid algorithms.

Popularity-based algorithms are simple and widely implemented – think of the *most-read* boxes on many news sites (Thorson, 2008). Collaborative filtering, in contrast, recommends to the individual what users who behave similarly have been reading previously (Garcin et al., 2012). The popularity-based algorithms do not require any user profile data. In contrast, collaborative filtering-based algorithms benefit from extensive user profiles. Semantic filtering-based algorithms rely on the similarity of content instead of the similarity of users (Capelle et al., 2012). While semantic filtering can provide more comprehensive information on a topic, it is also prone to delivering more of the same content. Finally, hybrid algorithms combine different indicators of news relevance for the users. Considering user and item similarities enables more nuanced modeling of user interests and can "balance prediction accuracy with other quality factors like novelty or diversity" (Karimi et al., 2018, p. 1206). A common approach for implementing hybrid algorithms combines collaborative and semantic filtering (Darvishy et al., 2020), but more complex hybrid approaches are possible (see the review by Karimi et al., 2018). As this short overview illustrates, it is moot to speculate about the dangers of echo chambers, filter bubbles, or privacy breaches, without considering the exact design of the tool in question.

While the implementation of recommender systems has been particularly relevant for journalism-related research, a growing number of studies are also looking at other algorithm-powered tools. One example is the design of systems for the automated generation of journalistic content – also known as algorithmic journalism (Anderson, 2013; Dörr, 2016).

## Algorithm Design and Content Production

Diakopoulos (2019) highlights the growing use of algorithms throughout the journalistic work chain: Automated data mining may be used to gather information and identify promising stories; text writing can be (partly) performed by algorithm-powered tools, and the distribution of content increasingly relies on automated social-media accounts (news bots; Lokot & Diakopoulos, 2016) or news recommenders (Karimi et al., 2018). The latter has frequently been the focus of discussions. A distribution tool can be created once and *left alone*. However, before the recent advent of accessible AI tools like ChatGPT, such technology would have been difficult for journalists to access.

The few exceptions were rather niche: Sports journalism has often been discussed as an example of the potential of automatizing the production of journalistic content (Kunert, 2020). Sports coverage is relatively easy to automate by filling in text templates with who scored a goal (or got a penalty) in which minute, which can be straightforward to provide in machine-readable form. The quality will often be worse than that of a dedicated sports journalist, but may nevertheless be sufficient (see Chapter 14). However, beyond such highly structured forms of reporting, the unpredictable nature of news made it harder to adopt automated content generation across the newsroom (Diakopoulos, 2019). This changed profoundly with the emergence of foundation models (Bommasani et al., 2021) and their application for developing new forms of generative AI. The so-called large language models (LLMs) excel at content creation tasks, which is probably best illustrated by the substantial interest generated by the release of ChatGPT, an LLM-powered chatbot developed by

OpenAI. While this may be the most widely known example, many similar LLM-powered applications can also be hosted on one's server. This enables the LLMs to be used for journalistic content production in multiple ways.

First, journalists can – just like any other person – use ChatGPT to produce content. This practice raises many ethical, professional, and legal concerns (Gondwe, 2023; Pavlik, 2023). However, given how widespread the use of such tools has become, it would be naive to assume that it is not at least occasionally used in journalism as well (Gondwe, 2023). Second, news organizations can directly integrate LLM-powered tools into their products via an application programming interface (API): The Norwegian public broadcaster NRK provides summaries of articles automatically generated by OpenAI. Third, news organizations could use alternative generative AI applications, publicly available or developed in-house, and build their systems around them.

It is important to realize that there is quite a difference between journalistic design decisions that can be compared here to recommender systems: While instructions to an LLM-powered tool for summarizing a text can be tweaked in various ways, ultimately, we are dealing with a black box. It differs from explicitly specifying a formula for ranking content for an algorithm powering a news recommender system.

## Algorithm Design and News Recommenders

News recommenders are one of the most apparent examples of incorporating journalistic values into algorithm design. The importance of news recommenders for journalistic practices prompted active discussions about how these algorithms can be designed (Helberger, 2021; Møller, 2023). In the related field of software design, questions regarding algorithm design often focus on the underlying logic for computational tools. The logic of a popularity-based news recommender algorithm is easy to implement – but it also raises normative concerns. These range from the ease of manipulating the outcome (see Chapter 16), by artificially generating clicks and likes (the so-called *inauthentic behavior*), to the questionable (at least implicitly) assumption that the most popular stories are also the most societally relevant.

Without going too much into technical details, one may say that designing an algorithm for journalistic purposes requires conscious decisions about not only what features are to be considered (what characteristics of a user and/or a news item should the algorithm take into account) and which mathematical model should be used to weigh them but also what to optimize for. Should a tool for news distribution prioritize what is clicked often with the potential aim of increasing user subscriptions? Or should it be optimized to inform citizens better (and, if yes, how can informedness be operationalized)? Or should something else be prioritized?

An increasing amount of work on value-sensitive algorithmic design moves beyond a simplistic, more-clicks-are-always-better logic. A recent literature review analyzed 183 papers on what is referred to as "beyond accuracy" values in news recommender systems (Bauer et al., 2024). Traditionally, recommender systems research focused on achieving high accuracy in predicting user choices: Given a dataset of user interactions, can we predict what a user would click on next (Bauer et al., 2024)? To some extent, this makes sense: Nobody wants a recommender system that recommends irrelevant news to a user. However, from a journalistic point of view, it is problematic to fully cater to audience demand without considering other values.

In contrast, value-driven design centers on the importance of systematically integrating different types of human values in developing computational systems (Friedman et al., 2013). Some examples of moral and professional values often discussed in the context of value-sensitive design include privacy, autonomy, trust, and courtesy; however, the exact selection of the values varies depends, among other things, on the sector for which the system is developed (Bastian et al. 2021; Friedman et al., 2013).

Two values that hold particular importance in the journalistic context, in particular news recommendation algorithms, are diversity and serendipity. Yet, it is challenging to operationalize these rather broad concepts into something usable for a specific algorithmic system. In the case of news diversity, there are multiple possible operationalizations, ranging from diversity of interpretations to diversity of sources (Joris et al., 2020); similarly, the way serendipity is defined can vary between particular models (Reviglio, 2019). Nevertheless, recent work on transferring these values in the algorithm design is promising. Vrijenhoek et al. (2022) presented a framework for incorporating normatively grounded diversity metrics into a news recommender algorithm design. Kiddle et al. (2023) proposed to design news recommender systems around "navigable surprise," which they see as finding a balance between recommending content relevant to the user yet surprising enough to allow for serendipitous encounters.

## Journalistic Values and Algorithm Design

In the past years, the research on the role and possible integration of journalistic perspectives in algorithm design has evolved rapidly, ranging from earlier works on the intersection of information technology and ethics (Ananny, 2016) to approaches to value-sensitive design for algorithm-powered tools (Bastian et al., 2021; Møller, 2023).

As journalism holds a special role in democratic societies, the ethical needs, requirements, and pitfalls of algorithms used there differ from other sectors (Helberger, 2021; Helberger et al., 2022). For instance, the issue of who is in charge of editorial control and at what point, is of paramount importance and requires accounting for the technologically changing environment that affects the creation and dissemination of news. The matters of editorial control and autonomy become essential in the context of sensitive issues for coverage, such as war, health crises, or climate change (Bastian et al., 2019; Canavilhas, 2022; Mach et al., 2021; Makhortykh & Bastian, 2022). Moreover, these issues are related to other sensitive aspects of the distribution of power on a societal level, including the digital divide or information equity (Díaz-Noci, 2023; see Chapter 23).

Traditionally, journalistic ethics debates focused on the behavior of individual journalists, which is shaped by the values of individuals themselves, their training and experiences, and the routines they follow in their daily work (see Chapter 18). However, in addition to this micro level of journalistic ethics (Ward, 2020), organizational structures play an important role (see Chapter 7). Examples of such structures are professional organizations providing ethical guidelines like codes of ethics or newsrooms establishing their own rules and codes of conduct or similar instruments to hold individual practitioners accountable (Fengler et al., 2022).

When algorithms come into play, journalistic ethics face more challenges and have to adapt to a new environment, some of which go hand in hand with broader anxieties regarding the use of big data (Richards & King, 2014). The same applies to regulatory frameworks, where changes happen slowly (Porlezza, 2023). Whereas traditionally, key actors in journalistic ethics were journalists, media organizations, and the public, today's journalism must account for private stakeholders with their own (profit-oriented) interests and goals. These stakeholders include nonjournalist individuals earning money from information dissemination, particularly social media influencers (Wellman et al., 2020) and powerful social media platforms (Nielsen & Ganter, 2022).

Transparency becomes a key issue for journalism, as society expects journalistic media not to act opaquely, in contrast to business entities. Here, media ethics and legal studies intersect and benefit from each other (Helberger et al., 2022; Porlezza, 2023). This is also due to changing requirements to specific roles inside newsrooms and/or the vanishing of some aspects of journalistic roles. Together, these technology-driven changes impact the possibilities and requirements for the journalistic profession as a whole (López Jiménez & Ouariachi, 2021; Mellado & Hermida, 2021) and the execution of journalistic values.

There is no consensus on a definite list of values that journalists and the media should follow. Still, depending on the political and economic context, the media system or journalistic culture, or the journalists' training, core values can be identified, which are crucial for journalism to fulfill its function in a democracy (see Chapters 14 and 18). Key concepts in this realm are public service, objectivity, autonomy, immediacy, and ethics (Deuze, 2005). On a more detailed level, the values of transparency, diversity, and credibility have played a crucial role. Following Bastian et al. (2021), differentiating between organization-centered and audience-centered values helps identify the core aspects of ethical handling of algorithm-powered tools in the journalistic field. However, identifying which groups of actors are supposed to promote certain values is paramount for deciding which values should be addressed when designing algorithm-powered tools.

In this context, it is useful to reiterate the following point from Helberger et al. (2022):

> identifying values and principles of responsible AI is an important starting point, nothing less and nothing more. Truly responsible use of journalistic AI is less about lists and more about the responsible organization of processes: the processes that result in the identification of relevant values, but also ways to concentrate, contest, formalize, implement, measure and continuously improve the way journalistic AI lives up to these values. (p. 1620)

The complexity of assessing the interactions between algorithm design and journalistic routines is attributed to the diversity of contexts in which these interactions occur. Today's newsrooms involve multiple journalistic roles and tasks, and many of them are profoundly influenced by algorithm-powered tools and, consequently, these tools' design. Often, adopting these tools aims to enhance journalists' efficiency, but the outcomes of this process can vary. Sometimes, it results in algorithms successfully undertaking routine tasks, which frees the time and resources of journalists and lets them produce higher-quality content through extensive research, but in other cases, it puts journalists' jobs at stake (Peña-Fernández et al., 2023; Portugal et al., 2023). A constructive approach is to help journalists develop new specific skills which are harder or impossible to automate (Portugal et al., 2023).

While algorithm-powered tools have a major impact on journalistic roles and routines, a significant challenge concerns the limited ability of individual journalists to influence these tools' design unless the feedback mechanisms are established by news organizations (see Section 4). In any event, journalists need to be aware of the functionality of the respective algorithm (e.g., how it decides what content to prioritize or what content it cannot produce) to assess what will happen to the content they produce or the information they research. While the traditional news production cycle, including research and dissemination, was largely transparent to the journalists, they may not be aware of all its aspects nowadays. Unlike fairly transparent popularity-based ranking algorithms, algorithm-powered tools for investigative journalism differ widely and result in different outcomes (Stray, 2021). Additionally, the workings of computational news discovery tools that notify journalists about potentially newsworthy content impact their traditional gatekeeping roles, often without making clear *how* the gatekeeping is done (Diakopoulos, 2020; see Chapter 11).

When news media increasingly employ algorithm-powered tools – and they do or are in the process of preparing to do so – journalists can no longer hide behind not being programmers or technical experts. While, indeed, in the case of earlier technological innovations, accountability may have been distributed between programmers, engineers, and journalists, this is not the case anymore. The technical details of how a website is made responsive, for example, by rendering properly on different devices, such as mobile and web browsers, might be something journalists can safely ignore and leave to others. However, this is not the case for news production and dissemination algorithms. Here, it is paramount for journalists to have at least a basic technical understanding of the respective tools.

Therefore, transparency, with proper algorithmic literacy, is a basic prerequisite of algorithm design in the journalistic sphere (see Chapter 12). Transparency is necessary to ensure that algorithm-powered tools used by journalists adhere to journalistic standards and do not undermine the role of journalism in democratic societies. It stresses that transparency is crucial not only for the tools and apps used by the audiences of specific media outlets but also equally important to journalists working at these outlets' newsrooms.

We need to stress again that not all algorithms are a black box. Some are transparent (think of an algorithm ranking content by popularity – you could do the same thing by hand, albeit less efficiently), and some can be made transparent through additional mechanisms, such as explanations of outcomes of specific algorithm-powered tools (Sullivan et al., 2019). From the perspective of journalistic ethics, this is very different from an LLM, where the inner workings are not easily understandable even for experts, and explanations would require substantial effort to be translated into a nonexpert language.

# Methodological Approaches to Study Algorithm Design in Journalism

The degree to which algorithm-powered tools are implemented in journalistic practice differs strongly between media systems and news organizations. While many organizations have run pilot projects, the role of algorithm-powered tools at most news organizations in Europe is currently limited. Outside of the Global North, the adoption of algorithmic innovations into journalism develops even slower due to limited resources often available to, in particular, local media organizations (Kothari & Cruikshank, 2022; Munoriyarwa et al., 2023; see Chapter 4). This results in a limited integration of journalistic approaches into algorithm design, especially in the Global South, further hindered by the lack of specific technical resources and the disconnectedness of local social processes (Gondwe, 2024; see Chapter 5).

However, there are exceptions, especially for the Global North: In Norway, since 2019, Schibsted, a major publisher, has used algorithmically ranked front pages – except for the first six items, which are manually selected (Borchgrevink-Brækhus, 2022). In this case, the ranking algorithm uses a combination of click-based metrics, conversion (sales of subscriptions), and news values (assigned by journalists) to create a newspaper's front page. This example illustrates that newsrooms actively try to find a balance between algorithmic curation and incorporating journalistic values, thus prompting the importance of understanding methodological approaches used for studying their relationship.

The existing research on the integration of journalistic perspectives into the algorithm design can be structured according to two lines: (1) methodology used to acquire information about how such integration happens (or should happen) and (2) specific aspects of journalistic perspectives and values, which are integrated into the algorithm design. In the case of the first line – i.e., methodology – it is possible to identify several approaches that prevail in the field of research. Because the journalistic field is currently discovering and grasping how algorithms can be designed to meet specific requirements that are important to fulfill journalistic functions, qualitative methods play a significant role in understanding these processes through a thorough investigation of available case studies.

It is also important to note that most research on integrating journalistic approaches into algorithm design currently focuses on the Global North (see Chapter 4). There are a few studies that have conducted interviews to examine how journalists outside of the Global North use algorithm-powered tools and more recent forms of generative AI (Gondwe, 2023;

Kothari & Cruikshank, 2022; Soto-Sanfiel et al., 2022; Trang et al., 2024). However, to our knowledge, none of them engaged specifically with integrating journalistic approaches into the design of these tools. Partially, it can be due to many journalistic organizations in the Global South (for African countries, see Kothari & Cruikshank, 2022), having limited resources and, consequently, having to rely on off-the-shelf tools developed in other regions, such as the Global North or China.

Several studies used in-depth interviews with practitioners to scrutinize their perspectives and strategies for integrating journalistic values in design. Mitova et al. (2023b) identified through interviews with Dutch and Swiss news organizations several approaches for keeping journalists in the algorithmic loop and introducing mechanisms for overriding or influencing how news recommenders weigh content. Bastian et al. (2021) looked at news organizations in the same two countries and found several difficulties in including journalists and their perspectives in the design process of the algorithmic system, specifically focusing again on news recommenders. In contrast, Diakopoulos (2020) relied on interviews to examine practitioners' perspectives on the design of algorithmic systems used for news discovery and also observed the importance of keeping the journalist in the loop.

Besides interviews, several studies also applied participatory and co-creation approaches. Portugal et al. (2023) employed a design thinking approach to examine how journalism students from Germany see possibilities for integrating journalistic standards into the development of AI applications for creating journalistic (video) content. Petridis et al. (2023) used a co-design approach to collect journalistic requirements for AI-powered creativity support tools.

Finally, several studies relied on ethnographic approaches. Gutierrez Lopez et al. (2023) applied design ethnography to study algorithmic design perceptions and practices at BBC. They identified several strategies for blending editorial values and technological capacities, from keeping editorial voices in the loop to integrating data into the newsroom culture (Gutierrez Lopez et al., 2023). Schjøtt Hansen and Hartley (2023) combined online and offline ethnography to examine the process of developing news personalization algorithms in a Denmark-based media organization and how the integration of journalistic principles is negotiated as part of its process.

The specific aspects of journalistic practices that must be integrated into the algorithmic design also vary. One key aspect is professional values: objectivity, diversity, editorial autonomy, and many more. Bastian et al. (2021) examined how 34 organization- and audience-centered values can be influenced by and incorporated into news recommender design. Gutierrez Lopez et al. (2023) noted the importance of integrating organizational values, albeit without an in-depth discussion of what these values imply. Other studies, including Diakopoulos (2020), implicitly discussed integrating individual values in design processes. Besides professional values, some studies occasionally considered other aspects of practices. Portugal et al. (2023) explored the possibility of integrating journalistic standards like comprehensibility or immediacy. Petridis et al. (2023) looked at concrete goals journalists can achieve using AI-driven tools, including facilitating trust or countering positive bias in the press.

All of this is not to say that research into algorithmic practices is limited to qualitative approaches. Ultimately, whether algorithms meet the requirements formulated in theory is an empirical question that often can be studied quantitatively. As an example, if a stated goal is that a recommendation system should not lead to information bubbles in which some users are unaware of some topics that the newsroom considers important, then it can be measured with the so-called algorithm audits or agent-based testing (Haim, 2020; Ulloa et al., 2024). One approach could be to set up automated bots that mimic the behavior of users on news websites or search engines to observe the algorithmically curated content they receive and assess its compliance with the public's and practitioners' expectations (Urman et al., 2021). For instance, Möller et al. (2018) tested how far using different recommendation algorithms impacts the diversity of recommended content from a Dutch newspaper. Several recent systematic reviews provide a comprehensive overview of algorithm audit research and possible insights for algorithm design (Bandy & Diakopoulos, 2020; Urman et al., 2024).

## Conclusion and Outlook

The general discourse surrounding the relationship between algorithm-powered tools and journalism nowadays is not primarily about *whether* journalistic organizations should use algorithmic tools but *how* they should apply them (Mitova et al., 2023a; Vrijenhoek et al., 2024). Talking to practitioners often shows that journalistic organizations are discussing algorithm-powered applications and are experimenting with pilot projects but often have fewer systems in production than one may think (Bastian et al., 2021).

Strikingly, this contrasts with public perceptions. Surveys in six countries revealed that most respondents think that journalists at least sometimes use AI, even for tasks like writing whole articles, next to less substantive applications like grammar checks (Fletcher & Nielsen, 2024). The percentage of people who believe that journalists do not use AI for various tasks is negligible (Fletcher & Nielsen, 2024). This implies that now is the time to work on integrating journalistic values into algorithm design. As more and more algorithm-powered tools will be deployed, ensuring that they align with journalistic approaches is crucial.

With algorithms increasingly shaping different aspects of the journalistic work cycle – from news production to news dissemination – it is of paramount importance to find ways to integrate core principles and values of journalism into the algorithm design. Part of this process involves considering the changing roles of journalists in the digital ecosystem and how these roles are influenced by new technologies (see Chapters 10 and 11). Examining the role of algorithms should go hand in hand with the role of journalists on the one hand and the public on the other hand. This way, integrating journalistic approaches into algorithm design becomes more effective, as it does not operate in a vacuum but in a specific context.

In particular, we propose that journalistic approaches to algorithm design should always consider three aspects: First, the features that the algorithm should take into account. These can be content features (e.g., topic or style), and the user features (e.g., interests or sociodemographics). The choice of which features should be included in the algorithm design must be based on normative considerations. Second, one needs to consider how the algorithm should weigh these features. Often, these will be automated, machine-learned weights, but it may be desirable to give journalists some influence here. A prime example is the possibility to override a ranking in case of breaking news. Third, consideration needs to be given to what the algorithm optimizes for. Traditionally, these are clicks – the more clicks, the better. However, there is a growing recognition that focusing solely on short-term goals may be detrimental to the long-term goals of a journalistic organization, like maintaining a good reputation (Bauer et al., 2024). Hence, an algorithm that only recommends what is more likely to be clicked on may not be the best option (Bauer et al., 2024). These three aspects, taken together, are important for bridging the design of algorithms with journalistic approaches.

When designing algorithms for journalism, it is important to consider that private media organizations generally operate at the intersection of pursuing economic interests and being accountable to the public and the state because of their (societal) functions in democracies and their role in media freedom (Vīķe-Freiberga et al., 2013). Therefore, algorithm designs should account for not only the political system, which defines the role that respective media organizations are expected to play in their societal context, but also the media system, which influences the business models of private, public, and state-owned media.

To sustain their existence, private media organizations have to follow the objective of earning money, and editors, reporters, data scientists, or administrative staff may have other priorities than investors or directors (see Chapters 7 and 8). Balancing public interests with

economic – and sometimes political – purposes is not new for media organizations. This struggle has accompanied the journalistic sector for many decades. However, it becomes particularly present when certain processes are automatized by an algorithm, and decisions regarding such automatization have to be made before certain issues occur as a result of it and not as a reaction.

An important component of the successful integration of journalistic approaches and algorithm design is the advancement of algorithmic and AI literacies among media practitioners (see Chapter 12). This is important not only for more productive and responsible use of algorithm-powered tools in journalism (Deuze & Beckett, 2022) but also for enabling a better comprehension of how journalistic practices can be translated into algorithmic design and whether certain aspects can be challenging to integrate.

Another direction for future research concerns how approaches specific to different journalism cultures and media systems can be integrated into the design of algorithm-powered tools used in newsrooms. While certain values and norms can be viewed as universal for journalism around the world (e.g., truth, accountability, and independence), a number of studies (Hanitzsch, 2007; Hanusch & Banjac, 2021) also highlight the presence of strong national and cultural differences in how these values are interpreted and adapted to practice (see Chapters 14 and 18). Moreover, in order to make sense of global ethics and shared values, the respective framework certainly has to be taken into account (Christians et al., 2009; Hanitzsch, 2007; Hanusch & Banjac, 2021; see Chapter 5). Under these conditions, it is essential to consider how these differences can affect algorithm designs and to avoid the suppression of regional differences and values.

Finally, another research direction relates to the advancement of generative AI and its adoption by journalists. As new models for text, image, and sound generation continue to emerge, it is vital to explore their potential to transform journalistic practices and to develop journalistic approaches that leverage these generative AI models and tools.

## Recommended References for Further Reading

Bauer, C., Bagchi, C., Hundogan, O., & van Es, K. (2024). Where are the values? A systematic literature review on news recommender systems. In *ACM Transactions on Recommender Systems*, *2*(3), 1–40. 10.1145/3654805

This review article summarized existing research on the role of different values in the context of news recommender design.

Diakopoulos, N. (2020). Computational news discovery: Towards design considerations for editorial orientation algorithms in journalism. *Digital Journalism, 8*(7), 945–967. 10.1080/21670811.2020.1736946

The article discusses approaches for prototyping tools for computational news discovery and their alignment with journalistic principles.

Kiddle, R., Welbers, K., Kroon, A., & Trilling, D. (2023). Enabling serendipitous news discovery experiences by designing for navigable surprise. *1st Workshop on Normative Design and Evaluation of Recommender Systems (NORMalize)*. Retrieved July 26, 2024, from https://ceur-ws.org/Vol-3639/short2.pdf

A paper showcasing the possibilities for integrating the value of serendipity into the algorithm design in the context of journalism.

References

Ananny, M. (2016). Toward an ethics of algorithms: Convening, observation, probability, and timeliness. *Science, Technology & Human Values*, *41*(1), 93–117. 10.1177/0162243915606523

Anderson, C. (2013). Towards a sociology of computational and algorithmic journalism. *New Media & Society*, *15*(7), 1005–1021. 10.1177/1461444812465137

Bandy, J., & Diakopoulos, N. (2020). Auditing news curation systems: A case study examining algorithmic and editorial logic in Apple News. *Proceedings of the International AAAI Conference on Web and Social Media*, *14(1)*, 36–47. 10.1609/icwsm.v14i1.7277

Bastian, M., Helberger, N., & Makhortykh, M. (2021). Safeguarding the journalistic DNA: Attitudes towards the role of professional values in algorithmic news recommender designs. *Digital Journalism*, *9*(6), 835–863. 10.1080/21670811.2021.1912622

Bastian, M., Makhortykh, M., & Dobber, T. (2019). News personalization for peace: How algorithmic recommendations can impact conflict coverage. *International Journal of Conflict Management*, *30*(3), 309–328. 10.1108/IJCMA-02-2019-0032

Bauer, C., Bagchi, C., Hundogan, O., & van Es, K. (2024). Where are the values? A systematic literature review on news recommender systems. In *ACM Transactions on Recommender Systems*, *2*(3), 1–40. 10.1145/3654805

Bommasani, R., Hudson, D. A., Adeli, E., Altman, R., Arora, S., von Arx, S., Bertstein, M.S., Bohg, J., Bosselut, A., Brunskill, E., Brynjolfsson, E., Buch, S., Card, D.,

Castellon, R., Chatterji, N., Chen, A., Creel, K., Davis, J.Q., Demszky, D., ... & Liang, P. (2021). *On the opportunities and risks of foundation models*. ArXiv. https://doi.org/10.48550/arXiv.2108.07258

Borchgrevink-Brækhus, M. (2022). «Det er ikke plass til alt på internett»: Algoritmestyrte forsider og redaksjonelle vurderinger. *Norsk medietidsskrift, 29*(3), 1–17. 10.18261/nmt.29.3.4

Canavilhas, J. (2022). Artificial intelligence in journalism: Automatic translation and recommendation system in the project "A European Perspective" (EBU). *Revista Latina de Comunicación Social, 80*, 1–13. 10.4185/RLCS-2022-1534

Capelle, M., Frasincar, F., Moerland, M., & Hogenboom, F. (2012). Semantics-based news recommendation. *WIMS '12: Proceedings of the 2nd International Conference on Web Intelligence, Mining and Semantics* (pp. 1–9). Association for Computing Machinery. https://doi.org/10.1145/2254129.2254163

Christians, C. G., Glasser, T. L., McQuail, D., Nordenstreng, K., & White, R. A. (2009). *Normative theories of the media: Journalism in democratic societies*. University of Illinois Press.

Darvishy, A., Ibrahim, H., Sidi, F., & Mustapha, A. (2020). HYPNER: A hybrid approach for personalized news recommendation. *IEEE Access, 8*, 46877–46894. 10.1109/ACCESS.2020.2978505

Deuze, M. (2005). What is journalism? Professional identity and ideology of journalists reconsidered. *Journalism, 6*(4), 442–464. 10.1177/1464884905056815

Deuze, M., & Beckett, C. (2022). Imagination, algorithms and news: Developing AI literacy for journalism. *Digital Journalism, 10*(10), 1913–1918. 10.1080/21670811.2022.2119152

Diakopoulos, N. (2019). *Automating the news: How algorithms are rewriting the media*. Harvard University Press.

Diakopoulos, N. (2020). Computational news discovery: Towards design considerations for editorial orientation algorithms in journalism. *Digital Journalism, 8*(7), 945–967. 10.1080/21670811.2020.1736946

Díaz-Noci, J. (2023). *Investigar la brecha digital, las noticias y los medios: Hacia la equidad informativa digital*. Ediciones Profesionales de la Información SL. https://doi.org/10.3145/digidoc-informe8

Dörr, K. N. (2016). Mapping the field of algorithmic journalism. *Digital Journalism, 4*(6), 700–722. 10.1080/21670811.2015.1096748

Dourish, P. (2016). Algorithms and their others: Algorithmic culture in context. *Big Data & Society*, *3*(2), 1–11. 10.1177/2053951716665128

Eskens, S., Helberger, N., & Möller, J. (2017). Challenged by news personalisation: Five perspectives on the right to receive information. *Journal of Media Law, 9*(2), 259–284. 10.1080/17577632.2017.1387353

Fengler, S., Eberwein, T., & Karmasin, M. (2022). *The global handbook of media accountability*. Routledge.

Fletcher, R., & Nielsen, R. (2024). *AI and the future of news: What does the public in six countries think of generative AI in news?* Reuters Institute for the Study of Journalism. https://doi.org/10.60625/RISJ-4ZB8-CG87

Friedman, B., Kahn, P., Borning, A., & Huldtgren, A. (2013). Value sensitive design and information systems. In N. Doorn, D. Schuurbiers, I. van de Poel & M. Gorman (Eds.), *Early engagement and new technologies: Opening up the laboratory* (pp. 55–95). Springer. https://doi.org/10.1007/978-94-007-7844-3_4

Garcin, F., Zhou, K., Faltings, B., & Schickel, V. (2012). Personalized news recommendation based on collaborative filtering. *2012 IEEE/WIC/ACM International Conferences on Web Intelligence and Intelligent Agent Technology* (pp. 437–441). IEEE. https://doi.org/10.1109/WI-IAT.2012.95

Gondwe, G. (2023). CHATGPT and the Global South: How are journalists in Sub-Saharan Africa engaging with generative AI? *Online Media and Global Communication, 2*(2), 228–249. 10.1515/omgc-2023-0023

Gondwe, G. (2024). Artificial intelligence, journalism, and the Ubuntu robot in Sub-Saharan Africa: Towards a normative framework. *Digital Journalism*, 1–19. 10.1080/21670811.2024.2311258

Gutierrez Lopez, M., Porlezza, C., Cooper, G., Makri, S., MacFarlane, A., & Missaoui, S. (2023). A question of design: Strategies for embedding AI-driven tools into journalistic work routines. *Digital Journalism, 11*(3), 484–503. 10.1080/21670811.2022.2043759

Haim, M. (2020). Agent-based testing: An automated approach toward artificial reactions to human behavior. *Journalism Studies, 21*(7), 895–911. 10.1080/1461670X.2019.1702892

Hanitzsch, T. (2007). Deconstructing journalism culture: Toward a universal theory. *Communication Theory, 17*(4), 367–385. 10.1111/j.1468-2885.2007.00303.x

Hanusch, F., & Banjac, S. (2021). Do journalists share universal values? In S. J. A. Ward (Ed.), *Handbook of global media ethics* (pp. 71–90). Springer. https://doi.org/10.1007/978-3-319-32103-5_5

Helberger, N., Eskens, S., van Drunen, M., Bastian, M., & Möller, J. (2020). *Implications of AI-driven tools in the media for freedom of expression*. Council of Europe.

Helberger, N. (2021). On the democratic role of news recommenders. In N. Thurman, S. Lewis, & J. Kunert (Eds.), *Algorithms, automation, and news* (pp. 14–33). Routledge.

Helberger, N., van Drunen, M., Möller, J., Vrijenhoek, S., & Eskens, S. (2022). Towards a normative perspective on journalistic AI: Embracing the messy reality of normative ideals. *Digital Journalism, 10*(10), 1605–1626. 10.1080/21670811.2022.2152195

Joris, G., De Grove, F., Van Damme, K., & De Marez, L. (2020). News diversity reconsidered: A systematic literature review unraveling the diversity in conceptualizations. *Journalism Studies, 21*(13), 1893–1912. 10.1080/1461670X.2020.1797527

Karimi, M., Jannach, D., & Jugovac, M. (2018). News recommender systems – survey and roads ahead. *Information Processing & Management, 54*(6), 1203–1227. 10.1016/j.ipm.2018.04.008

Kiddle, R., Welbers, K., Kroon, A., & Trilling, D. (2023). Enabling serendipitous news discovery experiences by designing for navigable surprise. *1st workshop on normative design and evaluation of recommender systems (NORMalize)*. Retrieved July 26, 2024, from https://ceur-ws.org/Vol-3639/short2.pdf

Kothari, A., & Cruikshank, S. A. (2022). Artificial intelligence and journalism: An agenda for journalism research in Africa. *African Journalism Studies*, *43*(1), 17–33. 10.1080/23743670.2021.1999840

Kunert, J. (2020). Automation in sports reporting: Strategies of data providers, software providers, and media outlets. *Media and Communication, 8*(3), 5–15. 10.17645/mac.v8i3.2996

Lokot, T., & Diakopoulos, N. (2016). News bots: Automating news and information dissemination on Twitter. *Digital Journalism, 4*(6), 682–699. 10.1080/21670811.2015.1081822

López Jiménez, E. A., & Ouariachi, T. (2021). An exploration of the impact of artificial intelligence (AI) and automation for communication professionals. *Journal of

Information, Communication and Ethics in Society, *19*(2), 249–267. 10.1108/JICES-03-2020-0034

Mach, K. J., Reyes, R. S., Pentz, B., Taylor, J., Costa, C. A., Cruz, S. G., Thomas, K. E., Arnott, J. C., Donald, R., Jagannathan, K., Kirchhoff, C. J., Rosella, L. C., & Klenk, N. (2021). News media coverage of covid-19 public health and policy information. *Humanities and Social Sciences Communications, 8*, 1–11. 10.1057/s41599-021-00900-z

Makhortykh, M., & Bastian, M. (2022). Personalizing the war: Perspectives for the adoption of news recommendation algorithms in the media coverage of the conflict in Eastern Ukraine. *Media, War & Conflict*, *15*(1), 25–45. 10.1177/1750635220906254

Mellado, C., & Hermida, A. (2021). The promoter, celebrity, and joker roles in journalists' social media performance. *Social Media+Society*, *7*(1), 1–11. 10.1177/2056305121990643

Mitova, E., Blassnig, S., Strikovic, E., Urman, A., de Vreese, C., & Esser, F. (2023a). Exploring users' desire for transparency and control in news recommender systems: A five-nation study. *Journalism*, 1–21. 10.1177/14648849231222099

Mitova, E., Blassnig, S., Strikovic, E., Urman, A., de Vreese, C., & Esser, F. (2023b). When worlds collide: Journalistic, market, and tech logics in the adoption of news recommender systems. *Journalism Studies*, *24*(16), 1957–1976. 10.1080/1461670X.2023.2260504

Møller, L. (2023). Designing algorithmic editors: How newspapers embed and encode journalistic values into news recommender systems. *Digital Journalism*, 1–19. 10.1080/21670811.2023.2215832

Möller, J., Trilling, D., Helberger, N., & van Es, B. (2018). Do not blame it on the algorithm: An empirical assessment of multiple recommender systems and their impact on content diversity. *Information, Communication & Society, 21*(7), 959–977. 10.1080/1369118X.2018.1444076

Munoriyarwa, A., Chiumbu, S., & Motsaathebe, G. (2023). Artificial intelligence practices in everyday news production: The case of South Africa's mainstream newsrooms. Journalism Practice, *17*(7), 1374–1392. 10.1080/17512786.2021.1984976

Nielsen, R., & Ganter, S. (2022). *The power of platforms: Shaping media and society*. Oxford University Press.

Pavlik, J. V. (2023). Collaborating with ChatGPT: Considering the implications of generative artificial intelligence for journalism and media education. *Journalism & Mass Communication Educator, 78*(1), 84–93. 10.1177/10776958221149577

Peña-Fernández, S., Meso-Ayerdi, K., Larrondo-Ureta, A., & Díaz-Noci, J. (2023). Without journalists, there is no journalism: The social dimension of generative artificial intelligence in the media. *Profesional de la información*, *32*(2), 1–15. 10.3145/epi.2023.mar.27

Petridis, S., Diakopoulos, N., Crowston, K., Hansen, M., Henderson, K., Jastrzebski, S., Nickerson, J., & Chilton, L. (2023). Anglekindling: Supporting journalistic angle ideation with large language models. *CHI '23: Proceedings of the 2023 CHI Conference on Human Factors in Computing Systems* (pp. 1–16). Association for Computing Machinery. https://doi.org/10.1145/3544548.3580907

Porlezza, C. (2023). Promoting responsible AI: A European perspective on the governance of artificial intelligence in media and journalism. *Communications*, *48*(3), 370–394. 10.1515/commun-2022-0091

Portugal, R., Wilczek, B., Eder, M., Thurman, N., & Haim, M. (2023). *Design thinking for journalism in the AI age: Towards an innovation process for responsible AI applications*. City Research Online. Retrieved July 26, 2024, from https://openaccess.city.ac.uk/id/eprint/30698/1/CJ_DataJConf_2023_paper_65%20%282%29.pdf

Reviglio, U. (2019). Serendipity as an emerging design principle of the infosphere: Challenges and opportunities. *Ethics and Information Technology*, *21*(2), 151–166. 10.1007/s10676-018-9496-y

Richards, N., & King, J. (2014). Big data ethics. *Wake Forest Law Review, 49*, 393–433.

Schjøtt Hansen, A., & Hartley, J. (2023). Designing what's news: An ethnography of a personalization algorithm and the data-driven (re) assembling of the news. *Digital Journalism, 11*(6), 924–942. 10.1080/21670811.2021.1988861

Soto-Sanfiel, M. T., Ibiti, A., Machado, M., Marín Ochoa, B. E., Mendoza Michilot, M., Rosell Arce, C. G., & Angulo-Brunet, A. (2022). In search of the Global South: assessing attitudes of Latin American journalists to artificial intelligence in journalism. *Journalism Studies*, *23*(10), 1197–1224. 10.1080/1461670X.2022.2075786

Stray, J. (2021). Making artificial intelligence work for investigative journalism. In N. Thurman, S. Lewis, & J. Kunert (Eds.), *Algorithms, automation, and news* (pp. 97–118). Routledge.

Sullivan, E., Bountouridis, D., Harambam, J., Najafian, S., Loecherbach, F., Makhortykh, M., Kelen, D., Wilkinson, D., Graus, D., & Tintarev, N. (2019). Reading news with a purpose: Explaining user profiles for self-actualization. *UMAP '19 Adjunct: Adjunct Publication of the 27th Conference on User Modeling, Adaptation and Personalization* (pp. 241–245). Association for Computing Machinery. https://doi.org/10.1145/3314183.3323456

Thorson, E. (2008). Changing patterns of news consumption and participation: News recommendation engines. *Information, Communication & Society*, *11*(4), 473–489. 10.1080/13691180801999027

Trang, T. T. N., Chien Thang, P., Hai, L. D., Phuong, V. T., & Quy, T. Q. (2024). Understanding the adoption of artificial intelligence in journalism: An empirical study in Vietnam. *SAGE Open*, *14*(2), 1–16. 10.1177/21582440241255241

Ulloa, R., Makhortykh, M., & Urman, A. (2024). Scaling up search engine audits: Practical insights for algorithm auditing. *Journal of Information Science*, *50*(2), 404–419. 10.1177/01655515221093029

Urman, A., Makhortykh, M., & Hannak, A. (2024). *Mapping the field of algorithm auditing: A systematic literature review identifying research trends, linguistic and geographical disparities*. arXiv. https://doi.org/10.48550/arXiv.2401.11194

Urman, A., Makhortykh, M., & Ulloa, R. (2021). Auditing source diversity bias in video search results using virtual agents. *WWW '21: Companion Proceedings of the Web Conference 2021* (pp. 232–236). Association for Computing Machinery. https://doi.org/10.1145/3442442.3452306

Vīķe-Freiberga, V., Däubler-Gmelin, H., Hammersley, B., & Maduro, L. M. P. P. (2013). *A free and pluralistic media to sustain European democracy.* High Level Group on Media Freedom and Media Pluralism. Retrieved July 26, 2024, from https://ec.europa.eu/information_society/media_taskforce/doc/pluralism/hlg/hlg_final_report.pdf

Vrijenhoek, S., Bénédict, G., Gutierrez Granada, M., Odijk, D., & De Rijke, M. (2022). RADio – rank-aware divergence metrics to measure normative diversity in news recommendations. *RecSys '22: Proceedings of the 16th ACM Conference* *on Recommender Systems* (pp. 208–219). Association for Computing Machinery. https://doi.org/10.1145/3523227.3546780

Vrijenhoek, S., Daniil, S., Sandel, J., & Hollink, L. (2024). Diversity of what? On the different conceptualizations of diversity in recommender systems. *FAccT '24: Proceedings of the 2024 ACM Conference on Fairness, Accountability, and Transparency* (pp. 573–584). Association for Computing Machinery. https://doi.org/10.1145/3630106.3658926

Ward, S. J. (2020). Journalism ethics. In K. Wahl-Jorgensen & T. Hanitzsch (Eds.), *The handbook of journalism studies* 2 pp. 307–323). Routledge.

Wellman, M., Stoldt, R., Tully, M., & Ekdale, B. (2020). Ethics of authenticity: Social media influencers and the production of sponsored content. *Journal of Media Ethics, 35*(2), 68–82. 10.1080/23736992.2020.1736078